\documentclass[aps,pra,longbibliography,twocolumn,showpacs,nofootinbib,superscriptaddress,notitlepage,10pt]{revtex4-2}
\usepackage{amsmath,amssymb,amsthm} % For advanced math typesetting
\usepackage{physics}
\usepackage{graphicx}% To include figures
\usepackage{hyperref}        % For clickable links and references
\usepackage{bm}   
 \usepackage{braket}   
\usepackage{subfig}

\usepackage{microtype} %微调字符宽度和标点突出部分
\usepackage{newtxtext}
\usepackage{tikz}
\usepackage{quantikz}

\hypersetup{
    colorlinks=true,
    linkcolor=blue,
    citecolor=blue,
    urlcolor=blue
}

\tikzset{
   circlecross/.style={
        draw, 
        circle, 
        minimum size=0pt, 
        inner sep=0.5pt, 
        fill=white,path picture={
            \draw[black] 
                (path picture bounding box.south) -- (path picture bounding box.north)
                (path picture bounding box.west) -- (path picture bounding box.east);
        }
        }
}
\tikzset{
    circlewc/.style={draw, circle, minimum size=0pt, inner sep=0.5pt, fill=white,   path picture={\draw[black] 
                (path picture bounding box.west) -- (path picture bounding box.east);
        }
        }
}
\newcommand{\htarg}{\gate[style={circlewc}]{}}
\newcommand{\ftarg}{\gate[style={circlecross}]{}}

\date{\today}
\begin{document}

\title{Optimized Gottesman-Kitaev-Preskill Error Correction via Tunable Preprocessing}

\author{Xiang-Jiang Chen}
\author{Hao-Miao Jiang}
\author{Liu-Jun Wang}
\email{ljwangq@ynu.edu.cn}
\author{Qing Chen}
\email{chenqing@ynu.edu.cn}
\affiliation{School of Physics and Astronomy and Yunnan Key Laboratory for Quantum Information, Yunnan University, Kunming 650500, China}

\begin{abstract}
 The Gottesman-Kitaev-Preskill (GKP) code is a promising bosonic candidate for realizing fault-tolerant quantum computation. Among existing error-correction protocols for GKP code, the Steane-type scheme is a canonical and widely adopted paradigm, yet its intrinsic noise propagation pattern limits further performance improvement. In this work, we propose a preprocessing-based Steane-type (P-Steane) scheme, which introduces a tunable preprocessing stage with squeezing parameters $a$ and $b$ to actively reshape noise propagation, thereby constituting a parameter framework. This framework spans a spectrum of protocols beyond existing methods, reproducing the performance of both the ME-Steane scheme ($a=1$, $b=1$) and the teleportation-based scheme ($a=1/\sqrt{2}$, $b=\sqrt{2}$) as special cases. Crucially, in the small-noise regime and when the data qubit is noisier than the ancilla qubits, P-Steane scheme achieves the minimum product of position- and momentum-quadrature output noise variances when $2a = b$, and consistently outperforms the ME-Steane scheme within a specific squeezing-parameter range under this condition. 
\end{abstract}
\maketitle

\section{Introduction}
The realization of fault-tolerant  quantum computation depends on effective quantum error correction (QEC) to protect fragile quantum states from decoherence~\cite{Shor1995scheme,gottesman1997stabilizer,terhal2015quantum}. Standard QEC methods typically encode logical information into multiple physical qubits, resulting in substantial hardware overhead that limits the scalability of quantum systems. Bosonic codes provide a more hardware-efficient alternative by encoding a logical qubit into a single bosonic mode~\cite{weedbrook2012gaussian,braunstein2005quantum}, leveraging its infinite-dimensional Hilbert space to provide the necessary redundancy. Representative bosonic codes include  cat code~\cite{ofek2016extending,leghtas2013hardware}, binomial code~\cite{hu2019quantum,michael2016new}, rotation-symmetric code~\cite{endo2025quantum,grimsmo2020quantum}, and Gottesman-Kitaev-Preskill (GKP) code~\cite{gottesman2001encoding,grimsmo2021quantum}. Among these codes, the GKP code stands out for its ability to efficiently correct small displacement errors in phase space~\cite{gottesman2001encoding} and photon loss~\cite{albert2018performance}. Although proposed over twenty years ago, its experimental realization has only recently been demonstrated on ion-trap~\cite{fluhmann2019encoding,de2022error,matsos2024robust,valahu2025quantum}, superconducting~\cite{campagne2020quantum,eickbusch2022fast,sivak2023real,lachance2024autonomous,brock2025quantum}, and photonic platforms~\cite{konno2024logical,larsen2025integrated}.

QEC for the GKP code relies on stabilizer measurements, which require coupling the encoded GKP state to auxiliary states. Two main approaches have been explored, distinguished by the type of auxiliary states employed~\cite{roy2025decoding}. The first approach performs phase estimation using a two-level ancilla~\cite{terhal2016encoding,royer2020stabilization}, but is limited by the fact that each measurement yields only a single bit of information, necessitating numerous repetitions to obtain the syndrome with high precision~\cite{grimsmo2021quantum}. The second approach uses fresh GKP states as auxiliaries, including the Steane-type~\cite{gottesman2001encoding,glancy2006error,wan2020memory} and teleportation-based schemes~\cite{walshe2020continuous,noh2022low}; the former is motivated by the Steane error-correction scheme for CSS codes~\cite{steane1997active}, whereas the latter follows Knill’s teleportation scheme~\cite{knill2005quantum}. Recently, a Steane-type scheme  with maximum likelihood estimation (ME-Steane)~\cite{fukui2023high} has attracted considerable attention, achieving improved performance compared with the original Steane scheme~\cite{zhang2021quantum,fukui2023high,wang2022multidimensional}. Furthermore, the ME-Steane and teleportation-based schemes display complementary behavior in the position and momentum quadratures~\cite{zhang2021quantum}, reflecting a trade-off in performance.

In GKP error-correction architectures, error propagation plays a crucial role in affecting performance~\cite{noh2022low}. For instance, performance can be improved by actively shaping error propagation, such as through asymmetric auxiliary state preparation~\cite{siegele2023robust} or appropriate logical gate selection~\cite{noh2020encoding}. The ME-Steane scheme retains the original Steane-type circuit and therefore preserves its intrinsic noise propagation pattern, which limits further performance improvement. To overcome this  performance limitation,  we propose a preprocessing-based Steane-type (P-Steane) scheme that introduces an active preprocessing stage into the original Steane-type circuit. In this stage, suitably tailored auxiliary states and logical gates—incorporating two tunable squeezing parameters $(a,b)$—are used to actively control noise propagation, thereby reshaping the output noise profile. Our analysis reveals that the P-Steane scheme serves as a unifying framework, encompassing the performance of both the ME-Steane and teleportation-based schemes as special cases, while its tunable parameters $(a,b)$ enable a wider range of noise-shaping strategies beyond these existing protocols. In a noise scenario where the small-noise regime holds and the data GKP qubit is noisier than the ancilla GKP qubits, the condition $2a = b$ minimizes the product of output noise variances in position and momentum quadratures. In particular, choosing the  parameters $(1/\sqrt{2}, \sqrt{2})$ renders P-Steane performance-equivalent to the teleportation-based scheme and yields symmetric output noise, while other choices produce asymmetric noise, enabling tailored optimization for different applications. Under this noise scenario and with $2a = b$, we further identify a squeezing-parameter range in which P-Steane consistently exhibits superior performance compared with the ME-Steane scheme.
 
The paper is organized as follows. In Sec.~\ref{sec-GKP states, operations, and noise model}, we introduce the fundamental properties of the GKP code and relevant logical gates, followed by a description of the Steane-type scheme and a brief comparison between the original Steane and ME-Steane schemes. In Sec.~\ref{sec-3}, we propose the P-Steane scheme and analyze its performance across various parameter settings. We then present numerical comparisons of the performance among four schemes: original Steane, ME-Steane, and two parameter settings of P-Steane scheme, one of which is performance-equivalent to teleportation-based scheme. Finally, Sec.~\ref{sec:4} concludes the paper with a discussion and outlook.

\section{PRELIMINARIES}\label{sec-GKP states, operations, and noise model}
\subsection{ The GKP error correction code }
Working in units where $\hbar = 1$, 
the displacement operators for the position and momentum quadratures are given by
\begin{eqnarray}
  \hat{X}(u) \equiv e^{-iu\hat{p}},\qquad
    \hat{Z}(v) \equiv e^{iv\hat{q}},
\end{eqnarray}
where $u, v \in \mathbb{R}$ denote the corresponding displacement amplitudes. 
A general displacement operator can be defined as \cite{stafford2023biased}
\begin{equation}\label{eq-general  displacement operator}
\hat{D}(u, v) \equiv \hat{X}(u)\hat{Z}(v)  = e^{-iu \hat{p}} e^{iv \hat{q}} .
\end{equation}
The code space of the ideal GKP code is stabilized by two commuting stabilizer generators
\begin{equation}
\hat{S_p}\equiv \hat{X}(2\sqrt{\pi})=e^{-i2\sqrt{\pi}\hat{p}},\quad
\hat{S_q}\equiv \hat{Z}(2\sqrt{\pi})=e^{i2\sqrt{\pi}\hat{q}}.
\end{equation}
The logical Pauli operators for the ideal GKP code are defined as
\begin{equation}
\bar{X} \equiv \hat{X}(\sqrt{\pi}) = e^{-i \sqrt{\pi} \hat{p}}, \quad \bar{Z} \equiv \hat{Z}(\sqrt{\pi}) = e^{i \sqrt{\pi} \hat{q}}.
\end{equation}
The operators $\bar{X}$ and $\bar{Z}$ commute with the stabilizers and generate them upon squaring, i.e., $\bar{X}^2 = \hat{S}_p$ and $\bar{Z}^2 = \hat{S}_q$. Therefore, the logical states in the computational basis (i.e., the eigenstates of $\bar{Z}$) are given by
\begin{equation}\label{eq-idea GKP 0 1}
\begin{aligned}
\ket{\bar{0}} &\propto \sum_{t \in \mathbb{Z}} \ket{q = 2t \sqrt{\pi}} \propto \sum_{n \in \mathbb{Z}} \ket{p = n\sqrt{\pi}}, \\
\ket{\bar{1}} &\propto \sum_{t \in \mathbb{Z}} \ket{q = (2t + 1)\sqrt{\pi}} \propto \sum_{n \in \mathbb{Z}} (-1)^n \ket{p = n\sqrt{\pi}}.
\end{aligned}
\end{equation}
Similarly, the logical states in the complementary basis (i.e., the eigenstates of $\bar{X}$) are given by
\begin{equation}\label{eq-idea GKP + -}
\begin{aligned}
\ket{\bar{+}} &\propto \sum_{t \in \mathbb{Z}} \ket{q = t \sqrt{\pi}} \propto \sum_{n \in \mathbb{Z}} \ket{p = 2n \sqrt{\pi}}, \\
\ket{\bar{-}} &\propto \sum_{t \in \mathbb{Z}} (-1)^t \ket{q = t\sqrt{\pi}} \propto \sum_{n \in \mathbb{Z}} \ket{p = (2n + 1)\sqrt{\pi}}.
\end{aligned}
\end{equation}

However, these ideal GKP states in Eqs.~(\ref{eq-idea GKP 0 1}) and (\ref{eq-idea GKP + -}) are unphysical, as their preparation would require infinite squeezing. In realistic scenarios, one must work with approximate GKP states of finite squeezing. An arbitrary approximate GKP state $\ket{\tilde{\psi}}$ can be regarded as the ideal GKP state $\ket{\bar{\psi}}$ subjected to coherent position and momentum shift errors with a Gaussian distribution~\cite{gottesman2001encoding,zhang2021quantum}, namely
\begin{equation}\label{eq-appr GKP}
\ket{\tilde{\psi}} = N \iint e^{-\frac{u^2+v^2}{2\Delta^2}} \hat{D}(u,v) \ket{\bar{\psi}} \, du \, dv,
\end{equation}
where $N$ is the normalization factor and $\hat{D}(u, v)$ is defined as in Eq.~\eqref{eq-general displacement operator}. The squeezing level is characterized by the Gaussian standard deviation $\Delta$, and $\ket{\tilde{\psi}}$ reduces to $\ket{\bar{\psi}}$ in the limit $\Delta \to 0$.

For efficient classical simulation, the coherent shift errors in approximate GKP states are commonly simplified to incoherent stochastic errors using the twirling approximation~\cite{noh2020fault}. In this work, we also employ the twirling approximation, assuming that shift errors arising from finite squeezing are incoherent.
Specifically, these errors can be modeled as resulting from a Gaussian shift error channel $\mathcal{E}$ with variance $\sigma^2$ acting on the ideal GKP state~\cite{zhang2021quantum}
\begin{equation}\label{eq-noise channel}
 \mathcal{E}(\ket{\bar{\psi}} \bra{\bar{\psi}})\! \equiv\! \!\iint \! \!P_\sigma(u) P_\sigma(v) \hat{D}(u,v) \ket{\bar{\psi}} \bra{\bar{\psi}} \hat{D}^\dagger(u,v)dudv,
\end{equation}
where $P_\sigma(x) = \frac{1}{\sqrt{2\pi \sigma^2}} e^{-\frac{x^2}{2\sigma^2}}$. For simplicity, the effect of the Gaussian shift error channel $\mathcal{E}$ acting on the ideal GKP state can also be expressed as
\begin{equation}\label{eq-new-form-gaussian-noise-channel}
\ket{\bar{\psi}} \xrightarrow{\mathcal{E}}  \hat{D}(u,v)\ket{\bar{\psi}},
\end{equation}
where $u,v$ are randomly drawn from an independent and identically distributed Gaussian distribution with zero mean and variance $\sigma^2$, i.e.,
$u,v \sim_{\text{IID}} \mathcal{N}(0, \sigma^2)$.
\subsection{Gaussian quantum gates }
In this section, we introduce several Gaussian quantum gates, along with the evolution of quadrature operators under their action.
We start by defining the single-mode squeezing gate as
\begin{equation}\label{eq-squeezing gate}
S(r) \equiv \exp\left( i \frac{\ln(r)}{2} (\hat{q}\hat{p} + \hat{p}\hat{q}) \right),
\end{equation}
where $r>0$ is the squeezing factor. In the Heisenberg picture, it transforms the quadrature operators as
\begin{equation}
\hat{q}\to  \hat{q}/r ,\quad \hat{p}\to  r \hat{p}.
\end{equation}
This evolution of quadrature operators can be represented by the circuit diagram 
\begin{equation}
\begin{quantikz}
\lstick{} &[-0.4cm]\wire[r][1]["\hat{q}"{above,pos=0.1},"\hat{p}"{below,pos=0.1}]{q} &\gate{S(r)}   \wire[r][1]["\hat{q}/r"{above,pos=0.6},"r\hat{p}"{below,pos=0.6}]{q} &.
 \end{quantikz}
\end{equation}
For $r > 1$, the squeezing gate suppresses position fluctuations while amplifying momentum fluctuations; the opposite holds for $0 < r < 1$.

Additionally, a two-mode SUM gate is defined as
\begin{equation}
\text{SUM}_{c\to t}\equiv \exp\!\left(-i\hat{q}_{c}\hat{p}_{t}\right).
\end{equation}
The SUM gate corresponds to the CNOT gate in discrete-variable systems, and it transforms the quadratures of the control (c) and target (t) modes as 
 \begin{equation}
 \begin{quantikz}[row sep=0.6cm]
 \lstick{} &\ctrl{1}\wire[l][1]["\hat{q}_c"{above,pos=0.5},"\hat{p}_c"{below,pos=0.5}]{q}  \wire[r][1]["\hat{q}_c"{above,pos=1.2},"\hat{p}_c-\hat{p}_t"{below,pos=1.2}]{q}&&\\
\lstick{}  & \ftarg \wire[l][1]["\hat{q}_t"{above,pos=0.5},"p_t"{below,pos=0.5}]{q}  \wire[r][1]["\hat{q}_t+\hat{q}_c"{above,pos=1.2},"\hat{p}_t"{below,pos=1.2}]{q} &&   
\end{quantikz}.
\end{equation}
Similarly, the Inverse-SUM gate is defined as
\begin{equation}
\text{SUM}_{c\to t}^{-1}\equiv \exp\!\left(i\hat{q}_{c}\hat{p}_{t}\right),
\end{equation}
and it transforms the quadratures of the control and target modes as 
 \begin{equation}
 \begin{quantikz}[row sep=0.6cm]
 \lstick{} &\ctrl{1}\wire[l][1]["\hat{q}_c"{above,pos=0.5},"\hat{p}_c"{below,pos=0.5}]{q}  \wire[r][1]["\hat{q}_c"{above,pos=1.2},"\hat{p}_c+\hat{p}_t"{below,pos=1.2}]{q}&&\\
\lstick{}  & \htarg \wire[l][1]["\hat{q}_t"{above,pos=0.5},"p_t"{below,pos=0.5}]{q}  \wire[r][1]["\hat{q}_t-\hat{q}_c"{above,pos=1.2},"\hat{p}_t"{below,pos=1.2}]{q} &&   
\end{quantikz}.
\end{equation}

Finally, the beam-splitter (BS) gate is defined as~\cite{noh2022low}
\begin{equation}
\begin{aligned}
BS_{j \to k}(\theta)
&\equiv \exp[-i\theta(\hat{q}_j \hat{p}_k - \hat{p}_j \hat{q}_k)]\\
&= \exp[\theta(\hat{a}^\dagger_j \hat{a}_k - \hat{a}_j \hat{a}^\dagger_k)].
\end{aligned}
\end{equation}
In particular, for a 50:50 BS gate ($\theta=\pi/4$), the corresponding quadrature transformations are given by
\begin{equation}
\begin{quantikz}[row sep=0.3cm]
  &\gate[2]{BS_{j\to k}}\wire[l][1]["\hat{q}_j"{above,pos=0.5},"\hat{p}_j"{below,pos=0.5}]{q}  \wire[r][1]["(\hat{q}_j-\hat{q}_k)/\sqrt{2}"{above,pos=1.7},"(\hat{p}_j-\hat{p}_k)/\sqrt{2}"{below,pos=1.7}]{q}&&&\\
  & \wire[l][1]["\hat{q}_k"{above,pos=0.5},"p_k"{below,pos=0.5}]{q}  \wire[r][1]["(\hat{q}_j+\hat{q}_k)/\sqrt{2}"{above,pos=1.7},"(\hat{p}_j+\hat{p}_k)/\sqrt{2}"{below,pos=1.7}]{q} &&&
 \end{quantikz}.
\end{equation}
\subsection{ Steane-type error correction}
\begin{figure}[tb] 
	\centering
\includegraphics[width=0.9\linewidth]{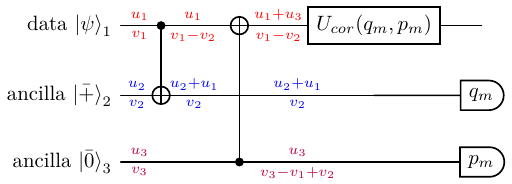}
	\caption{Steane-type error-correction circuit. The data qubit is coupled to two ancilla qubits via the $\mathrm{SUM}_{1\to 2}$ and $\mathrm{SUM}_{3\to 1}$ gates. A feedback correction shift $U_{\mathrm{cor}}(q_m,p_m)$ is then applied based on the homodyne detection outcomes $q_m$ and $p_m$, with the original Steane and ME-Steane schemes differing in the values of this shift. The labels ($u_i,v_i$) next to each rail indicate a general displacement error $\hat{D}_i(u_i,v_i)$, and the diagram shows how these errors propagate through the circuit.}
	\label{fig-Steane type circuit}
\end{figure}

In this section, we introduce two Steane-type schemes, including the original Steane~\cite{gottesman2001encoding} and the ME-Steane~\cite{zhang2021quantum,fukui2023high} schemes, and provide a brief comparison between them. The circuit of the Steane-type error correction scheme is illustrated in Fig.~\ref{fig-Steane type circuit}. The input state is given by
\begin{equation}\label{eq-q and p quadratures input state}
\hat{D}_1(u_1,v_1) \ket{\bar{\psi}}_1 \otimes \hat{D}_2(u_2,v_2) \ket{\bar{+}}_2 \otimes \hat{D}_3(u_3,v_3) \ket{\bar{0}}_3.
\end{equation}
Here, we assume that the shift errors of data qubit satisfy $u_1, v_1 \sim_{\text{IID}} \mathcal{N}(0, \sigma_D^2)$, while shift errors of the ancilla qubits satisfy $u_2, v_2, u_3, v_3 \sim_{\text{IID}} \mathcal{N}(0, \sigma_A^2)$. After applying the $\text{SUM}_{1 \to 2}$ and $\text{SUM}_{3 \to 1}$ gates, the output state (up to a global phase) is given by
\begin{equation}
\begin{aligned}
&\hat{D}_1(u_1+u_3,v_1-v_2) \ket{\bar{\psi}}_1  \otimes \hat{D}_2(u_2+u_1,v_2)  \ket{\bar{+}}_2  \\
&\otimes  \hat{D}_3(u_3 ,v_3-v_1+v_2) \ket{\bar{0}}_3   \\
\propto\,&  \hat{D}_1(u_1+u_3,v_1-v_2) \ket{\bar{\psi}}_1 \otimes \sum_{t,n \in \mathbb{Z}}\ket{q_2=t\sqrt{\pi}+u_2+u_1}\\
& \otimes \ket{p_3=n\sqrt{\pi}+v_3-v_1+v_2}.
\end{aligned}
\end{equation}
The measurement outcomes of the ancilla qubits are $q_m = t\sqrt{\pi} + u_2 + u_1$ and $p_m = n\sqrt{\pi} + v_3 - v_1 + v_2$, where $t,n \in \mathbb{Z}$.

Under the assumption that smaller shifts are more likely to happen than larger shifts, the measurement outcome $q_m$ (and $p_m$) deviates only slightly from the nearest $t\sqrt{\pi}$ (and $n\sqrt{\pi}$) with high probability. We define a function $R_{\sqrt{\pi}}(z)$ that gives the distance between $z$ and its nearest integer multiple of $\sqrt{\pi}$:
\begin{equation}\label{eq-def-R}
R_{\sqrt{\pi}}(z) \equiv z - n \sqrt{\pi},\,\text{for} \,\, (n - \frac{1}{2}) \sqrt{\pi} \leq z < (n + \frac{1}{2}) \sqrt{\pi}.
\end{equation}
Based on the measurement outcomes $q_m$ and $p_m$, a correction shift $U_{\mathrm{cor}}(q_m,p_m)$ is subsequently applied. In the original Steane scheme~\cite{gottesman2001encoding}, this correction shift is chosen as $\hat{D}_1 \left(-R_{\sqrt{\pi}}(q_m),-R_{\sqrt{\pi}}(-p_m)\right)$. Consequently, the output shifts of the data qubit in the $q$ and $p$ quadratures are given by
\begin{equation}\label{eq-original Steane output noise}
\begin{aligned}
 u_O'&=u_1+u_3- R_{\sqrt{\pi}}(u_1+u_2),\\
 v_O'&=v_1-v_2-R_{\sqrt{\pi}}(v_1-v_2-v_3).
\end{aligned}
\end{equation}

Notably, the data shift errors and the measured shift errors are correlated. Unlike the original Steane scheme, the ME-Steane scheme exploits this correlation, employing maximum-likelihood estimation to infer the actual data errors~\cite{fukui2023high}. Previous studies~\cite{zhang2021quantum,fukui2023high,wang2022multidimensional} have shown that the ME-Steane scheme outperforms the original Steane scheme, in which the $q$-quadrature correction was analyzed in detail; here, we extend the analysis to the correction of both $q$ and $p$ quadratures. Specifically, the correction shift applied in ME-Steane scheme is $\hat{D}_1(-\eta_q R_{\sqrt{\pi}}(q_m), -\eta_p R_{\sqrt{\pi}}(-p_m))$, where the scaling factors $\eta_q$ and $\eta_p$ are determined via maximum-likelihood estimation  and are given by (see  Appendix \ref{Appendix-ME- Scaling Factor Selection} for details)
\begin{equation}\label{eq-ME-Steane scaling factor selection}
\eta_q= \frac{\sigma_D^2}{\sigma_D^2 + \sigma_A^2},\quad \eta_p=\frac{\sigma_D^2 + \sigma_A^2}{\sigma_D^2 + 2 \sigma_A^2}.
\end{equation}
Consequently, the output shifts of the data qubit in the $q$ and $p$ quadratures are expressed as
\begin{equation}\label{eq-ME-Steane's output noise}
\begin{aligned}
 u_M'&=u_1+u_3-\eta_q R_{\sqrt{\pi}}(u_1+u_2),\\
 v_M'&=v_1-v_2-\eta_p R_{\sqrt{\pi}}(v_1-v_2-v_3).
\end{aligned}
\end{equation}

Here we present a brief comparison between the original Steane and ME-Steane schemes, while a rigorous numerical comparison is provided in Section~\ref{sec-Numerical Analysis}. Considering the case where small shift errors happen, with $|u_1+u_2|<\sqrt{\pi}/2$ and $|v_1-v_2-v_3|<\sqrt{\pi}/2$,  the output shifts for these two schemes (Eqs.~(\ref{eq-original Steane output noise}) and (\ref{eq-ME-Steane's output noise})) reduce to
\begin{equation}
\begin{aligned}
    u_O' &\to u_3 - u_2 \sim \mathcal{N}(0, \sigma_{O,q}^2),  \\
    v_O' &\to v_3 \sim \mathcal{N}(0, \sigma_{O,p}^2), \\
    u_M' &\to (1-\eta_q)u_1 - \eta_q u_2 + u_3 \sim \mathcal{N}(0, \sigma_{M,q}^2),  \\
    v_M' &\to (1-\eta_p)(v_1 - v_2) + \eta_p v_3 \sim \mathcal{N}(0, \sigma_{M,p}^2),
\end{aligned}
\end{equation}
where corresponding variances are given by
\begin{equation}\label{eq-OS and ME variance}
\begin{aligned}
    \sigma_{O,q}^2 &= 2\sigma_A^2, & \sigma_{O,p}^2 &= \sigma_A^2, \\
    \sigma_{M,q}^2 &= \frac{\sigma_A^2 (\sigma_A^2 + 2 \sigma_D^2)}{\sigma_A^2 + \sigma_D^2}, & \sigma_{M,p}^2 &= \frac{\sigma_A^2 (\sigma_A^2 + \sigma_D^2)}{2\sigma_A^2 + \sigma_D^2}.
\end{aligned}
\end{equation}
When $\sigma_A \ll \sigma_D$, we have $\eta_q, \eta_p \to 1$, and the ME-Steane scheme reduces to the original Steane scheme~\cite{zhang2021quantum}. In contrast, when $\sigma_A$ is comparable to $\sigma_D$, the output shifts in the ME-Steane scheme follow distributions with smaller variances, with $\sigma_{M,q}^2<\sigma_{O,q}^2$ and $\sigma_{M,p}^2<\sigma_{O,p}^2$, leading to superior performance.

\section{GKP error correction via tunable preprocessing}\label{sec-3}
 \subsection{P-Steane scheme} 
\begin{figure*}[!htb]
	\centering
\includegraphics[width=0.6\linewidth]{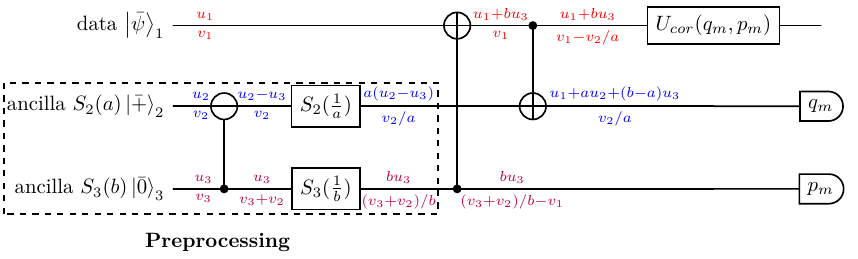}
	\caption{The circuit of the P-Steane scheme, comprising a preprocessing stage followed by a Steane-type correction stage. Labels ($u_i,v_i$) denote displacement errors, and the diagram illustrates their propagation through the circuit.}
	\label{fig-P-Steane circuit}
\end{figure*}
In this section, we propose a preprocessing-based Steane (P-Steane) scheme, which actively controls noise propagation and enhances performance by appropriately tailoring the ancilla inputs and logical gates. The circuit of the P-Steane error-correction scheme, shown in Fig.~\ref{fig-P-Steane circuit}, consists of a preprocessing stage followed by a Steane-type error-correction stage. To analyze the preprocessing stage, we first consider ideal auxiliary GKP states ($u_2=v_2=u_3=v_3=0$) for simplicity.  In the ideal case, the input auxiliary state is
\begin{equation}
\ket{\phi}_I=  S_2(a)\ket{\bar{+}}_2 \otimes S_3(b)\ket{\bar{0}}_3,  
\end{equation}
where $S_2(a)$ and $S_3(b)$ denote the squeezing operations defined in Eq.~(\ref{eq-squeezing gate}), with $a$ and $b$ as tunable squeezing parameters. Note that both $S_2(a)\ket{\bar{+}}_2$ and $S_3(b)\ket{\bar{0}}_3$ are rectangular-lattice GKP states~\cite{stafford2023biased,zhang2023concatenation}. The input state $\ket{\phi}_I$ is subsequently processed by the $\text{SUM}_{3\to 2}^{-1}$ gate, and then the squeezing gates $S_2(1/a)$ and $S_3(1/b)$. When the tunable squeezing parameters satisfy $2a/b \in\mathbb{Z}$, the output state after the preprocessing stage is given by
\begin{equation}\label{eq-ideal ancilla preprocessing}
\ket{\Phi}_I=U_1 \ket{\phi}_I =\ket{\bar{+}}_2 \ket{\bar{0}}_3 ,
\end{equation}
where $U_1 = S_2(1/a)S_3(1/b) \text{SUM}_{3\to 2}^{-1}$ is the unitary operator corresponding to the preprocessing operations. We now prove Eq.~(\ref{eq-ideal ancilla preprocessing}) using the stabilizer formalism.
\begin{proof}
$\ket{\bar{+}}_2  \ket{\bar{0}}_3$ is the unique quantum state (up to a global phase) which is stabilized by the set of stabilizer generators $S$
\begin{equation}
\begin{aligned}
S:\lbrace &S_1=e^{-i\sqrt{\pi}\hat{p}_2},S_2=e^{i2\sqrt{\pi} \hat{q}_2},\\
&S_3=e^{-i2\sqrt{\pi} \hat{p}_3},\ S_4=e^{i\sqrt{\pi} \hat{q}_3}\rbrace.
\end{aligned}
\end{equation}
Given that
\begin{equation}
\ket{\Phi}_I = U_1 \ket{\phi}_I = \left[ U_1 S_2(a) S_3(b) \right] \ket{\bar{+}}_2 \ket{\bar{0}}_3 = U_2 \ket{\bar{+}}_2 \ket{\bar{0}}_3,
\end{equation}
where $U_2 = U_1 S_2(a) S_3(b)$. Therefore, the state $\ket{\Phi}_I$ is stabilized by the transformed set of stabilizer generators $S' = U_2 S U_2^\dagger$, i.e.,
\begin{equation}
\begin{aligned}
S':\lbrace &S_1'=e^{-i\sqrt{\pi}\hat{p}_2},\ S_2'=e^{i2\sqrt{\pi} (\hat{q}_2+\frac{a}{b}\hat{q}_3)},\\
&S_3'=e^{-i2\sqrt{\pi} (\hat{p}_3 -\frac{a}{b}\hat{p}_2)},\ S_4'=e^{i\sqrt{\pi} \hat{q}_3}\rbrace.
\end{aligned}
\end{equation}
We define $m \equiv \frac{2a}{b} \in \mathbb{Z}$, yielding $S_2 = S_2'/ (S_4')^m$ and $S_3 = S_3'\times (S_1')^m$,
so the stabilizer groups generated by $S$ and $S'$ are identical. Hence, we have $\ket{\Phi}_I = \ket{\bar{+}}_2 \ket{\bar{0}}_3$.
\end{proof}

We now extend the analysis to noisy auxiliary GKP states, where the corresponding input is 
\begin{equation} \ket{\phi}_N = \hat{D}_2(u_2,v_2) \otimes \hat{D}_3(u_3,v_3) \ket{\phi}_I,\end{equation}
where the shift errors of the ancilla qubits are assumed to satisfy $u_2, v_2, u_3, v_3 \sim_{\text{IID}} \mathcal{N}(0, \sigma_A^2)$.
Within the preprocessing stage, these shift errors are transformed as follows (up to a global phase): 
\begin{equation}\label{eq-error propagation relations}
\begin{aligned}
&U_1 \left [\hat{D}_2(u_2,v_2) \otimes \hat{D}_3(u_3,v_3)\right ] U_1^{\dagger} \\
\to & \hat{D}_2\left(a(u_2-u_3), \frac{v_2}{a}\right) \otimes \hat{D}_3\left(b u_3, \frac{v_3 +v_2}{b}\right) .
\end{aligned}
\end{equation}
Combining this result with Eq.~(\ref{eq-ideal ancilla preprocessing}), one can find that when $m = 2a/b \in \mathbb{Z}$, the output state after preprocessing stage is given by
\begin{equation} 
\begin{aligned}
\ket{\Phi}_N=&U_1 \ket{\phi}_N  \\
 =&\hat{D}_2\left(a(u_2-u_3), \frac{v_2}{a}\right) \ket{\bar{+}}_2 \otimes \hat{D}_3\left(b u_3, \frac{v_3 + v_2}{b}\right)  \ket{\bar{0}}_3 .
\end{aligned}
\end{equation}

After completing the preprocessing stage, the P-Steane scheme proceeds with Steane-type error correction, with the order of the two SUM gates swapped. This swap only interchanges the output noise profiles along the $q$ and $p$ quadratures~\cite{zhang2021quantum}, without affecting overall performance of the scheme. This swap is adopted to facilitate a more intuitive comparison in the subsequent analysis. Under the P-Steane scheme, the output shifts of the data qubit in the $q$ and $p$ quadratures  are given by
\begin{equation}\label{eq-Output Noise under P-Steane}
\begin{aligned}
u_P'&=\xi_q-c_q\,R_{\sqrt{\pi}}(m_q)=\xi_q-c_qm_q-c_q n_q \sqrt{\pi} ,
\\v_P'&=\xi_p-c_p\,R_{\sqrt{\pi}}(m_p)=\xi_p-c_p m_p-c_p n_p \sqrt{\pi} ,
\end{aligned}
\end{equation}
where $\xi_q$ and $\xi_p$ are the data shift errors to be corrected, and $m_q$ and $m_p$ are the measurement shift errors, along the $q$ and $p$ quadratures, respectively. They are expressed as
\begin{equation}\label{eq-P-Steane-def}
\begin{pmatrix}
\xi_q \\ m_q \\ \xi_p \\ m_p
\end{pmatrix}
=
\begin{pmatrix}
u_1 + b u_3 \\
u_1 + \frac{bm}{2} u_2 + \frac{2b-bm}{2} u_3 \\
v_1 - \frac{2v_2}{bm} \\
v_1 - \frac{v_2+v_3}{b}
\end{pmatrix},
\end{equation}
where $
m \equiv \frac{2a}{b}\in\mathbb{Z}$. The scaling factors $c_q$ and $c_p$, determined via maximum-likelihood estimation method, are given by (see Appendix \ref{Appendix-ME- Scaling Factor Selection} for details)
\begin{equation}\label{eq-P-Steanescaling factor selection}
c_q=\frac{2 \sigma_D^2 + b^2 (2-m ) \sigma_A^2}{2 \sigma_D^2 + b^2 (m^2 - 2 m + 2) \sigma_A^2},\,c_p= \frac{b^2 m \sigma_D^2 + 2 \sigma_A^2}{b^2 m \sigma_D^2 + 2 m \sigma_A^2}.
\end{equation}
The integers $n_q$ and $n_p$ indicate whether a logical error occurs. If the measurement shift errors are less than $\frac{\sqrt{\pi}}{2}$, namely $|m_q|<\frac{\sqrt{\pi}}{2}$ and $|m_p|<\frac{\sqrt{\pi}}{2}$, we have $(n_q,n_p)=(0,0)$ and no logical error occurs; otherwise, a logical error  $\bar{X}^{n_q}\bar{Z}^{n_p}$ is induced for $(n_q, n_p) \neq (0,0)$.

\subsection{Parameter Analysis}\label{sec-Parameter Analysis}
Based on the output shifts of the P-Steane scheme given in Eq.~(\ref{eq-Output Noise under P-Steane}), we analyze its performance for different parameter combinations $(b, m)$. We first consider the parameter choice $b=1, m=2$ (thus $a=1$). In this case, the squeezing operators $S(a)$ and $S(b)$ reduce to the identity operator $I$, and Eq.~(\ref{eq-Output Noise under P-Steane}) simplifies to
\begin{equation}\label{P-Steane Output Noise, b=1, m=2}
\begin {aligned}
u_{P1}' &= u_1 + u_3 - \frac{\sigma_D^2}{\sigma_D^2 + \sigma_A^2} R_{\sqrt{\pi}} (u_1 + u_2), \\
v_{P1}' &= v_1 - v_2 - \frac{\sigma_D^2 + \sigma_A^2}{\sigma_D^2 + 2\sigma_A^2} R_{\sqrt{\pi}} (v_1 - v_2 - v_3).
\end{aligned}
\end{equation}
A comparison between Eq.~(\ref{eq-ME-Steane's output noise}) and Eq.~(\ref{P-Steane Output Noise, b=1, m=2}) shows that $u_{P1}' = u_{M}'$ and $v_{P1}' = v_{M}'$. Therefore, when choosing $b=1$ and $m=2$, the P-Steane and ME-Steane schemes exhibit identical performance.

A natural question then arises: is there a more favorable parameter combination $(b, m)$ that can further improve the performance? We start by considering the small-noise regime, where both $\sigma_D$ and $\sigma_A$ are small. In this regime, the measurement shift errors are less than $\sqrt{\pi}/2$, leading to $(n_q, n_p) = (0, 0)$. Consequently, the output shifts of the P-Steane scheme in Eq.~(\ref{eq-Output Noise under P-Steane}) simplifies to
\begin{equation}
\begin{aligned}
u_P' &\to \xi_q -c_q m_q  \sim \mathcal{N}(0,\sigma_q^2), \\
v_P'& \to \xi_p-c_p m_p \sim \mathcal{N}(0,\sigma_p^2),
\end{aligned}
\end{equation}
where the noise variances $\sigma_q^2$ and $\sigma_p^2$ quantify the performance in the small-noise regime and are given by
\begin{equation}\label{eq-P-Steane variance}
\begin{aligned}
\sigma_q^2&=\frac{\sigma_A^2\, m^2 (b^4 \sigma_A^2 + 2 b^2  \sigma_D^2)}{2 b^2 (m^2- 2 m+2) \sigma_A^2 + 4 \sigma_D^2},\\
\sigma_p^2&=\frac{4 \sigma_A^4 + 2 b^2 (2 - 2 m + m^2) \sigma_A^2 \sigma_D^2}{b^2 m^2 (2 \sigma_A^2 + b^2 \sigma_D^2)}.
\end{aligned}
\end{equation}
The product of these two variances can be expressed as
\begin{equation}\label{eq-product variance}
\begin{aligned}
\sigma_q^2 \sigma_p^2=&\sigma_A^4 \times \\
&[1+ 
\frac{ 2 b^2 (m-1)^2 ( \sigma_D^4-\sigma_A^4 )}{\left(b^2 ( m^2 - 2 m+2) \sigma_A^2 + 2 \sigma_D^2\right) (2 \sigma_A^2 + b^2 \sigma_D^2)} ].
 \end{aligned}
\end{equation}
Because error correction is performed using fresh auxiliary GKP states~\cite{gottesman2001encoding}, the data qubit is typically noisier than the ancilla qubits (i.e., $\sigma_D > \sigma_A$). In this work, we consider the more general case $\sigma_D \ge \sigma_A$, encompassing the special case in which the data and ancilla qubits have equal noise levels (i.e., $\sigma_D = \sigma_A$). Eq.~(\ref{eq-product variance}) shows that  $\sigma_q^2 \sigma_p^2$ attains its minimum value $\sigma_A^4$ in three cases: (i) $b=0$, (ii) $\sigma_D = \sigma_A$, and (iii) $m=1$. For the case of $b=0$, the situation corresponds to infinite squeezing, which is not physically realizable. When $\sigma_D = \sigma_A$, the minimum variance product is achieved for any choice of $(m, b)$. Finally, the case of $m=1$ is of particular interest,  as it yields the minimum value for all $\sigma_D \ge \sigma_A$. 

We now focus on the case of $m=1$ (i.e., $2a=b$), under which the P-Steane output shifts in Eq.~(\ref{eq-Output Noise under P-Steane}) and the variances in Eq.~(\ref{eq-P-Steane variance}) simplify, respectively, to
\begin{equation}\label{eq-output-m=1}
\begin{aligned}
u_{P2}' &= u_1 + b u_3 - R_{\sqrt{\pi}} \left( u_1 + \frac{b}{2}(u_2 + u_3) \right), \\
v_{P2}' &= v_1 - \frac{2v_2}{b} - R_{\sqrt{\pi}} \left( v_1 - \frac{v_2+v_3}{b} \right),
\end{aligned}
\end{equation}
and
\begin{equation}
\sigma_{q2}^2=\frac{b^2 \sigma_A^2}{2} ,\quad \sigma_{p2}^2=\frac{2\sigma_A^2}{b^2}.
\end{equation}

Setting $b = \sqrt{2}$ (thus $a=1/\sqrt{2}$) yields symmetric variances in the $q$ and $p$ quadratures:
\begin{equation}\label{eq-sy}
\sigma_{q2}^2 \to \sigma_{qS}^2 =\sigma_A^2, \quad \sigma_{p2}^2 \to \sigma_{pS}^2 =\sigma_A^2.
\end{equation}
Interestingly, in this symmetric case, the P-Steane scheme achieves performance identical to that of the teleportation-based scheme, as proven in Appendix~\ref{sec-equivalence-proof-of-P-Steane-and-Teleportation-based-schemes}. Comparing Eq.~\eqref{eq-sy} with Eq.~\eqref{eq-OS and ME variance}, one finds
\begin{equation}
\sigma_{qS}^2<\sigma_{M,q}^2,\quad \sigma_{pS}^2>\sigma_{M,p}^2.
\end{equation}
Thus, a trade-off arises in the small-noise regime: while the P-Steane scheme with $b = \sqrt{2}$ (along with the performance-equivalent teleportation-based scheme) excels in the $q$ quadrature, the ME-Steane scheme is superior in the $p$ quadrature.

We next consider a more meaningful scenario under which the P-Steane scheme consistently matches or surpasses the ME-Steane scheme in both quadratures within the small-noise regime, namely,
\begin{equation}\label{eq-simultaneous-optimality}
\sigma_{q2}^2 \le \sigma_{M,q}^2, \qquad 
\sigma_{p2}^2 \le \sigma_{M,p}^2 .
\end{equation}
We define a noise ratio $k \equiv \sigma_D^2/\sigma_A^2 \ge 1$, which quantifies the relative noise level between the data qubit and the ancilla qubits. By expressing $\sigma_{M,q}^2$ and $\sigma_{M,p}^2$ (see Eq.~\eqref{eq-OS and ME variance}) in terms of $k$,
Eq.~\eqref{eq-simultaneous-optimality} can be rewritten as a constraint on $b$:
\begin{equation}\label{eq-b-constraints}
\sqrt{3-\frac{k-1}{k+1}} \le b \le \sqrt{3+\frac{k-1}{k+1}} .
\end{equation}
For $k=1$, the inequality has a unique solution $b=\sqrt{3}$, yielding
$\sigma_{q2}^2 = \sigma_{M,q}^2$ and $\sigma_{p2}^2 =\sigma_{M,p}^2$. Importantly, for $k>1$, the solution set for $b$ forms a continuous interval throughout which the P-Steane scheme consistently outperforms the ME-Steane scheme in small-noise regime. 

We further consider the large-noise regime, where $\sigma_D$ and $\sigma_A$ are both large. In this regime, the measurement shift errors $m_q$ and $m_p$ may exceed $\sqrt{\pi}/2$, leading to $(n_q, n_p) \neq (0,0)$ and consequently introducing a Pauli error $\bar{X}^{n_q}\bar{Z}^{n_p}$. The probability
of having such a Pauli error is closely related to the variances of  $m_q$ and $m_p$. For $m=1$, the measurement shift errors in the P-Steane scheme (see Eq.~(\ref{eq-output-m=1})) follow the distributions:
\begin{equation}
\begin{aligned}
m_q = u_1 + \frac{b}{2}(u_2 + u_3) &\sim \mathcal{N}\left(0,\Sigma_{q}^2=\sigma_D^2+\frac{b^2}{2}\sigma_A^2\right), \\
m_p =  v_1 - \frac{v_2+v_3}{b} &\sim \mathcal{N}\left(0,\Sigma_{p}^2=\sigma_D^2+\frac{2}{b^2}\sigma_A^2\right).
\end{aligned}
\end{equation}
In contrast, the measurement shift errors of the ME-Steane scheme (see Eq.~\eqref{eq-ME-Steane's output noise}) follow the distributions:
\begin{equation}
\begin{aligned}
u_1+u_2 &\sim \mathcal{N}(0,\Sigma_{M,q}^2=\sigma_D^2+\sigma_A^2), \\
v_1-v_2-v_3& \sim \mathcal{N}(0,\Sigma_{M,p}^2=\sigma_D^2+2\sigma_A^2) .
\end{aligned}
\end{equation}
Within the range specified in Eq.~(\ref{eq-b-constraints}),
$\Sigma_q^2 >\Sigma_{M,q}^2$ and $\Sigma_p^2 < \Sigma_{M,p}^2$ always hold. This indicates that, in the large-noise regime, the P-Steane scheme has an advantage in the $p$ quadrature but a disadvantage in the $q$ quadrature.

\subsection{Numerical Results}\label{sec-Numerical Analysis}
\begin{figure*}
    \centering

    \subfloat[$k=\sigma_D^2/\sigma_A^2=1$]{
        \includegraphics[width=0.8\textwidth]{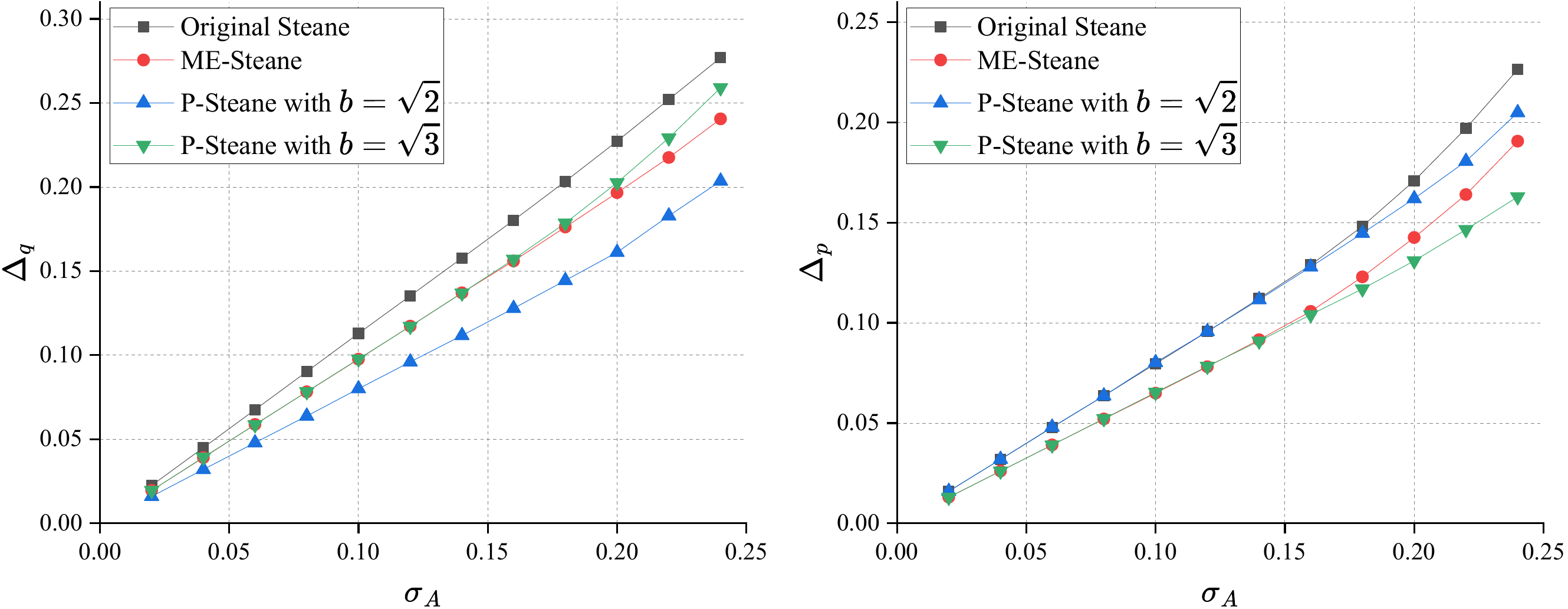}
        \label{fig-k1}
    }

    \vfill

    \subfloat[$k=\sigma_D^2/\sigma_A^2=3$]{
        \includegraphics[width=0.8\textwidth]{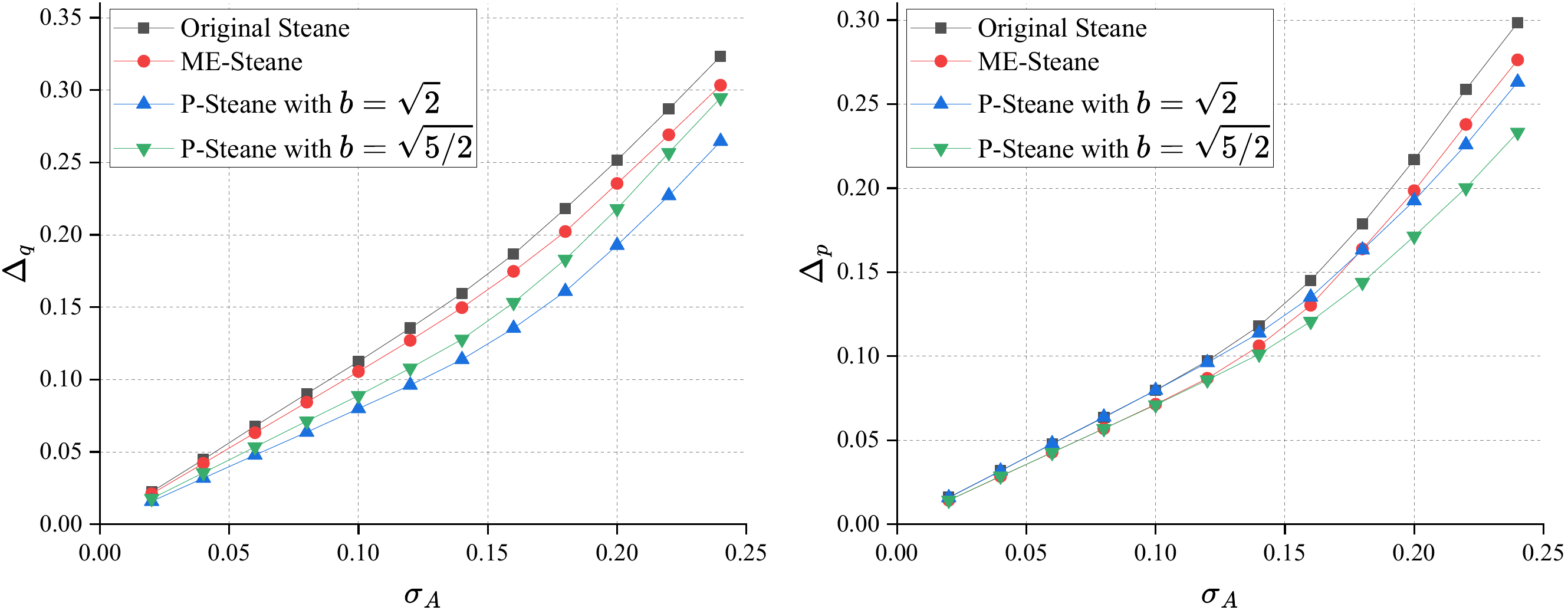}
        \label{fig-k3}
    }

    \caption{Comparison of the performance of four GKP error-correction schemes under two noise ratios: (a) $k = \sigma_D^2 / \sigma_A^2 = 1$ and (b) $k = 3$. The left and right columns show the performance metrics $\Delta_q$ and $\Delta_p$ as functions of the ancilla noise standard deviation $\sigma_A$, respectively. The schemes compared include the original Steane, the ME-Steane, and the P-Steane scheme with two parameter settings.}
    \label{fig-performance}
\end{figure*}

In this section, we present a numerical comparison of the performance of several GKP error-correction schemes. To quantify performance, we adopt the metric proposed in Ref.~\cite{zhang2021quantum}, which measures the deviation between a ideal GKP error correction and a noisy GKP error correction scheme. For convenience, we begin by defining the performance metric in the $q$ quadrature. In a ideal scenario, where the ancilla qubits are perfect with shift errors satisfying $u_2 = u_3 = 0$, the output shifts of data qubit after correction are then given by
\begin{equation}
u_I' = u_1 - R_{\sqrt{\pi}}(u_1).
\end{equation}
In a realistic scenario, the ancilla qubits are noisy with shift errors satisfying $u_2, u_3 \sim_{\mathrm{IID}} \mathcal{N}(0, \sigma_A^2)$, and we denote the output shift after error correction by $u_N'$. For the original Steane, ME-Steane, and P-Steane schemes, the corresponding expressions of $u_N'$ are given in Eqs.~(\ref{eq-original Steane output noise}), (\ref{eq-ME-Steane's output noise}), and (\ref{eq-Output Noise under P-Steane}), respectively. The performance metric for the $q$ quadrature is defined as~\cite{zhang2021quantum}
\begin{equation}\label{eq-Delta_q}
\Delta_q \equiv  \left\langle |(u_N' - u_I')\!\! \mod\! 2\sqrt{\pi}  |\right\rangle,
\end{equation} 
where the angle brackets denote the Gaussian-weighted average over the Gaussian variables $u_1$, $u_2$ and $u_3$, and the modulo $2\sqrt{\pi}$ accounts for the fact that shifts by integer multiples of $2\sqrt{\pi}$ do not cause logical errors for the GKP states. Similarly, for the $p$ quadrature, let $v_I'$ and $v_N'$ denote the output shifts in the ideal and realistic scenarios, respectively. The performance metric is then defined as
\begin{equation}\label{eq-Delta_p}
\Delta_p \equiv
\left\langle \bigl|(v_N' - v_I') \bmod 2\sqrt{\pi}\bigr| \right\rangle.
\end{equation}
Smaller values of $\Delta_q$ or $\Delta_p$ indicate better error-correction performance.

Fig.~\ref{fig-performance} shows the performance metrics $\Delta_q$ and $\Delta_p$ as functions of the noise standard deviation $\sigma_A$ for four error-correction schemes: the original Steane, ME-Steane, and two P-Steane schemes. Figs.~\ref{fig-k1} and~\ref{fig-k3} present the results for noise ratios $k=1$ and $k=3$, respectively.  For these two P-Steane schemes, we fix $m=1$ and set the squeezing parameter $b$ to either $\sqrt{2}$ or $\sqrt{3-(k-1)/(k+1)}$, where the latter expression yields $b = \sqrt{3}$ for $k=1$ and $b = \sqrt{5/2}$ for $k=3$. Motivated by experimentally realized 9.5 dB squeezing GKP states (corresponding to a noise standard deviation $\sigma \approx 0.236$) reported in Ref.~\cite{campagne2020quantum}, we restrict $\sigma_A$ to the range $0 < \sigma_A < 0.25$.

In the limit $\sigma_A \to 0$, all schemes approach ideal error correction, so both $\Delta_q$ and $\Delta_p$ approach zero. As $\sigma_A$ increases, $\Delta_q$ and $\Delta_p$ rise monotonically. For any nonzero $\sigma_A$, the original Steane scheme exhibits the largest $\Delta_q$ and $\Delta_p$, indicating the worst performance among these schemes. Therefore, we focus subsequent comparison on the three remaining schemes.

For the case of $k = 1$ (see Fig.~\ref{fig-k1}), the P-Steane scheme with $b=\sqrt{2}$ is optimal in the $q$ quadrature, whereas  P-Steane scheme with $b=\sqrt{3}$ performs best in the $p$ quadrature. A comparison between the ME-Steane scheme and the P-Steane scheme with $b=\sqrt{3}$ shows that their performances converge in the regime $\sigma_A < 0.16$, owing to the output noise variances satisfying $\sigma_{q2}^2 = \sigma_{M,q}^2$ and $\sigma_{p2}^2 = \sigma_{M,p}^2$. When $\sigma_A > 0.16$, their performances begin to diverge: the ME-Steane scheme outperforms in the $q$ quadrature but underperforms in the $p$ quadrature.
This divergence arises because the P-Steane scheme (with $b=\sqrt{3}$) exhibits a larger measurement noise variance in the $q$ quadrature ($\Sigma_{q2}^2 > \Sigma_{M,q}^2$), resulting in a higher probability of logical errors, whereas the situation is reversed in the $p$ quadrature.

For the case of $k=3$ (see Fig.~\ref{fig-k3}), the P-Steane scheme with $b=\sqrt{2}$ performs optimally in the $q$ quadrature, while the P-Steane scheme with $b=\sqrt{5/2}$ yields the best performance in the $p$ quadrature. Notably, when $\sigma_A < 0.25$, the P-Steane scheme with $b=\sqrt{5/2}$ consistently outperforms the ME-Steane scheme in both quadratures, demonstrating a significant advantage. In the $q$ quadrature, this advantage arises from the smaller output variance of the P-Steane scheme ($\sigma_{q2}^2 < \sigma_{M,q}^2$); 
in the $p$ quadrature, although the output noise variance is identical ($\sigma_{p2}^2 = \sigma_{M,p}^2$), the P-Steane scheme benefits from a lower probability of logical errors  due to the smaller measurement noise variance ($\Sigma_{p2}^2 < \Sigma_{M,p}^2$).

\section{CONCLUSION and DISCUSSION}
\label{sec:4}
In this work, we propose a preprocessing-based Steane-type (P-Steane) scheme that enhances error-correction performance by introducing an active preprocessing stage into the original Steane-type circuit. The P-Steane scheme reproduces the performance of the ME-Steane and teleportation-based schemes at the parameter settings $(a, b) = (1, 1)$ and $(1/\sqrt{2}, \sqrt{2})$, respectively, thereby establishing a unified framework connecting these two prominent schemes. Analyzing the output noise variances in the $q$ and $p$ quadratures ($\sigma_q^2$ and $\sigma_p^2$) within the small-noise regime, we find that under the condition $2a = b$, the variance product $\sigma_q^2 \sigma_p^2$ attains its minimum value $\sigma_A^4$ for all noise ratios $k \equiv \sigma_D^2 / \sigma_A^2 \ge 1$. Numerical results demonstrate that the P-Steane scheme exhibits a significant advantage under current experimental state-preparation conditions, for example with 9.5 dB squeezing GKP states (corresponding to a noise standard deviation $\sigma \approx 0.236$) reported in Ref.~\cite{campagne2020quantum}. For illustration, in the case of $k=3$, the P-Steane scheme with $(a,b)=(\sqrt{5/8},\,\sqrt{5/2})$ consistently outperforms the ME-Steane scheme over $0<\sigma_A<0.25$. Note that the noise model we consider assumes finite squeezing of the GKP states as the only source of noise, while extending it to include imperfections in logical gates and measurements would provide a more realistic characterization, which we leave for future work.

Although this work focuses on single-qubit GKP error correction, the P-Steane scheme can be naturally extended to broader applications owing to its ability to actively tailor the output noise profile. More specifically, in the P-Steane scheme, $(a,b)=(1/\sqrt{2},\sqrt{2})$ produces symmetric output noise, whereas other parameter choices satisfying $2a=b$ yield asymmetric noise. Here we discuss two scenarios in which such asymmetric noise profiles are advantageous. The first example is the concatenation of GKP codes with qubit codes tailored for asymmetric noise, such as the XZZX surface code~\cite{bonilla2021xzzx}, which exhibits a high threshold exceeding the hashing bound~\cite{bennett1996mixed} under asymmetric noise. In this setting, shaping the GKP output noise to align with the qubit code’s correctable noise profile is expected to enhance overall performance. Another example is the implementation of the non-Clifford $T$ gate on GKP states, which is an essential resource for universal quantum computation. A direct unitary implementation of the $T$ gate via the cubic phase gate suffers from an intrinsic logical error floor under symmetric noise, but this limitation can be removed under a asymmetric noise profile~\cite{hastrup2021unsuitability,nguyen2025fault}, which can be generated on demand by the P-Steane scheme.

~\\\\ 
{\bf Acknowledgements:} 
This work was supported by the National Natural Science Foundation of China (Grant Nos. 12165020 and 12464064), the Major Science and Technology Project of Yunnan Province (Grant No. 202502AD080015), and the Yunnan
Fundamental Research Project (Grant No. 202401AT070448).

\appendix
\vspace{0.5cm}

\begin{widetext}
\section{Proof of Eqs.~(\ref{eq-ME-Steane scaling factor selection}) and (\ref{eq-P-Steanescaling factor selection})} \label{Appendix-ME- Scaling Factor Selection}
For the ME-Steane scheme, let $\xi_q'$ and $\xi_p'$ be the data shift errors to be corrected, and $m_q'$ and $m_p'$ the measurement shift errors, along the $q$ and $p$ quadratures, respectively. They are given by (see Eq.~\eqref{eq-ME-Steane's output noise}):
\begin{equation}
    \begin{aligned}
\xi_q' &= u_1 + u_3 \sim \mathcal{N}(0,\sigma_D^2+\sigma_A^2), \\
 m_q'   &= u_1 + u_2 \sim \mathcal{N}(0,\sigma_D^2+\sigma_A^2), \\
\xi_p' &= v_1 - v_2 \sim \mathcal{N}(0,\sigma_D^2+\sigma_A^2), \\
 m_p'   &= v_1 -v_2-v_3\sim \mathcal{N}(0,\sigma_D^2+2\sigma_A^2).
    \end{aligned}
\end{equation}
Based on the measurement shift errors $m_q'$ and $m_p'$, we estimate the actual data shift errors as $\bar{\xi_q'}=\eta_q m_q'$ and $\bar{\xi_p'}=\eta_p m_p'$, where the coefficients $\eta_q$ and $\eta_p$ are chosen to minimize the estimation variances $\text{Var}(\xi_q' - \bar{\xi_q'})$ and $\text{Var}(\xi_p' - \bar{\xi_p'})$, respectively. For instance, in the $q$ quadrature, we have 
\begin{equation} \text{Var}(\xi_q'-\bar{\xi_q'})=\text{Var}(\xi_q'- \eta_q m_q')=\text{Var}(\xi_q')-2\eta_q\text{Cov}(\xi_q',m_q')+\eta_q^2\text{Var}(m_q'). \end{equation} 
This variance is minimized by choosing 
\begin{equation} \eta_q=\frac{\text{Cov}(\xi_q',m_q')}{\text{Var}(m_q')}= \frac{\sigma_D^2}{\sigma_D^2 + \sigma_A^2}. \end{equation} 
Similarly, in the $p$ quadrature, the optimal coefficient is
\begin{equation} \eta_p=\frac{\text{Cov}(\xi_p',m_p')}{\text{Var}(m_p')}=\frac{\sigma_D^2 + \sigma_A^2}{\sigma_D^2 + 2 \sigma_A^2}. \end{equation} Thus, Eq.~(\ref{eq-ME-Steane scaling factor selection}) is proved.

We analyze the P-Steane scheme in a manner analogous to the ME-Steane scheme.  We define $\xi_q$ ($\xi_p$) as the shift error to be corrected on the data qubit, and $m_q$ ($m_p$) as the measurement shift error in the $q$ ($p$) quadrature, respectively. According to Eq.~\eqref{eq-P-Steane-def}, they are given by
\begin{equation}
    \begin{aligned}
 \xi_q &= u_1 + b u_3 \sim \mathcal{N}(0,\sigma_D^2+b^2 \sigma_A^2) , \\
  m_q   &= u_1 + \frac{bm}{2} u_2 + \frac{2b-bm}{2} u_3  \sim \mathcal{N}(0,\sigma_D^2 + \frac{b^2(m^2 - 2m + 2)}{2} \sigma_A^2), \\
 \xi_p &= v_1 - \frac{2v_2}{bm} \sim \mathcal{N}(0,\sigma_D^2 + \frac{4}{b^2 m^2} \sigma_A^2), \\
 m_p   &= v_1 - \frac{v_2+v_3}{b} \sim \mathcal{N}(0,\sigma_D^2 + \frac{2}{b^2} \sigma_A^2).
     \end{aligned}
 \end{equation}
We estimate the actual data shift errors as $\bar{\xi}_q=c_q m_q$ and $\bar{\xi}_p=c_p m_p$. The coefficients $c_q$ and $c_p$ are chosen to minimize $\mathrm{Var}(\xi_q-\bar{\xi}_q)$ and $\mathrm{Var}(\xi_p-\bar{\xi}_p)$, respectively. Consequently, the optimal scaling factors are given by
\begin{equation}
\begin{aligned}
c_q&=\frac{\mathrm{Cov}(\xi_q,m_q)}{\mathrm{Var}(m_q)}
= \frac{2 \sigma_D^2 + b^2 (2-m ) \sigma_A^2}{2 \sigma_D^2 + b^2 (m^2 - 2 m + 2) \sigma_A^2},\
c_p&=\frac{\mathrm{Cov}(\xi_p,m_p)}{\mathrm{Var}(m_p)}
= \frac{b^2 m \sigma_D^2 + 2 \sigma_A^2}{b^2 m \sigma_D^2 + 2 m \sigma_A^2}.
\end{aligned}
\end{equation}

\section{Performance Equivalence Between the P-Steane Scheme with (m=1, b=\texorpdfstring{$\sqrt{2}$}{sqrt(2)}) and Teleportation-Based Scheme}
\label{sec-equivalence-proof-of-P-Steane-and-Teleportation-based-schemes}
When $m=1$ and $b=\sqrt{2}$, the output shifts of the P-Steane scheme (see Eq.~\eqref{eq-Output Noise under P-Steane}) simplify to
\begin{equation}\label{eq-ps}
 \begin{aligned} u_{P3}' &= u_1 + \sqrt{2} u_3 - R_{\sqrt{\pi}} \left( u_1 + \frac{u_2 + u_3}{\sqrt{2}} \right), \quad \ v_{P3}' &= v_1 - \sqrt{2} v_2 - R_{\sqrt{\pi}} \left( v_1 - \frac{v_2 + v_3}{\sqrt{2}} \right) .\end{aligned} \end{equation}
For the teleportation-based scheme~\cite{walshe2020continuous,noh2022low}, whose circuit is shown in Fig.~\ref{fig-teleportation-based circuit}, the encoded state is teleported from mode 1 to mode 3. A Pauli correction $\bar{P} \in \{\bar{I},\bar{X},\bar{Z},\bar{X}\bar{Z}\}$ is then applied to mode 3 based on the measurement outcomes $\sqrt{2} q_m$ and $\sqrt{2} p_m$~\cite{grimsmo2020quantum,noh2022low}. Defining a function
\begin{equation}\label{eq-pi(z)}
\pi(z)=
\begin{cases}
0, & \text{if } |z \!\!\! \mod 2\sqrt{\pi}| < \frac{\sqrt{\pi}}{2} \\
\sqrt{\pi} & \text{otherwise},
\end{cases},
\end{equation}
and the output shifts of the teleportation-based scheme can be written as
\begin{equation}\label{eq-appen-tele-output}
u_T=\frac{u_2+u_3}{\sqrt{2}} +\pi (u_1 -\frac{u_2-u_3}{\sqrt{2}}) ,\quad
v_T=\frac{v_2+v_3}{\sqrt{2}} +\pi(v_1 +\frac{v_2-v_3}{\sqrt{2}}).
\end{equation}
Using the definition of $R_{\sqrt{\pi}}(z)$ in Eq.~\eqref{eq-def-R}, we then obtain
\begin{equation}\label{eq-z-R}
z-R_{\sqrt{\pi}}(z) = 
\begin{cases}
    2n\sqrt{\pi} & \text{if } |z \!\!\! \mod 2\sqrt{\pi}| < \frac{\sqrt{\pi}}{2} \\
    (2n+1)\sqrt{\pi} & \text{if } \text{otherwise}
\end{cases} ,\text{for}\,\, (n - \frac{1}{2}) \sqrt{\pi} \leq z < (n + \frac{1}{2}) \sqrt{\pi}.
\end{equation}
Comparing Eqs.~\eqref{eq-pi(z)} and \eqref{eq-z-R}, one can find that the functions $\pi(z)$ and $z-R_{\sqrt{\pi}}(z)$ differ by a $2n\sqrt{\pi}$ shift. Such a shift does not affect the performance metrics $\Delta_q$ and $\Delta_p$, which include a modulo $2\sqrt{\pi}$ operation (see Eqs.~\eqref{eq-Delta_q} and \eqref{eq-Delta_p}). Therefore, the following output shifts are performance-equivalent to $u_T$ and $v_T$ in Eq.~\eqref{eq-appen-tele-output}:
\begin{equation}\label{eq-u_T',v_T'}
\begin{aligned}
u_T'&=\frac{u_2+u_3}{\sqrt{2}} +\left [(u_1 -\frac{u_2-u_3}{\sqrt{2}}) -R_{\sqrt{\pi}}(u_1 -\frac{u_2-u_3}{\sqrt{2}}) \right ]
 =u_1+\sqrt{2}u_3-R_{\sqrt{\pi}}(u_1 -\frac{u_2-u_3}{\sqrt{2}})\\
 v_T'&=\frac{v_2+v_3}{\sqrt{2}} +\left [ (v_1 +\frac{v_2-v_3}{\sqrt{2}})-R_{\sqrt{\pi}}(v_1 +\frac{v_2-v_3}{\sqrt{2}}) \right ]
=v_1+\sqrt{2}v_2 -R_{\sqrt{\pi}}(v_1 +\frac{v_2-v_3}{\sqrt{2}})
\end{aligned}
 \end{equation}

 \begin{figure}[htb]
	\centering
	\includegraphics[width=0.5\linewidth]{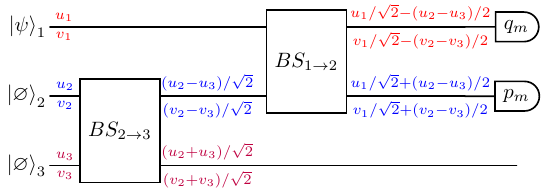}
	\caption{The teleportation-based error-correction circuit~\cite{walshe2020continuous,noh2022low} teleports the encoded state from mode 1 to mode 3, after which a Pauli correction $\bar{P} \in \{\bar{I},\bar{X},\bar{Z},\bar{X}\bar{Z}\}$ is applied to mode 3 based on the measurement outcomes $\sqrt{2} q_m$ and $\sqrt{2} p_m$. Labels ($u_i,v_i$) denote displacement errors, and the diagram illustrates their propagation through the circuit.}
 \label{fig-teleportation-based circuit}
\end{figure}

We now compute the probability density functions of the output shifts in Eqs.~\eqref{eq-ps} and \eqref{eq-u_T',v_T'}. In this work, the shift errors on the data qubit are modeled as $u_1, v_1 \sim_{\text{IID}} \mathcal{N}(0, \sigma_D^2)$, while those on the ancilla qubits are modeled as $u_2, v_2, u_3, v_3 \sim_{\text{IID}} \mathcal{N}(0, \sigma_A^2)$. As an example, we calculate the probability distribution of $u_{P3}'$ in Eq.~\eqref{eq-ps}. Define $X$ as the shift error to be corrected on the data qubit and $Y$ as the measurement shift error, namely
\begin{equation}
X \equiv u_1 + \sqrt{2} u_3 \sim \mathcal{N}(0,\sigma_{X}^2=\sigma_D^2+2\sigma_A^2),\quad Y \equiv  u_1 + \frac{u_2 + u_3}{\sqrt{2}} \sim \mathcal{N}(0,\sigma_{Y}^2=\sigma_D^2+\sigma_A^2).
\end{equation}
The correlation coefficient between $X$ and $Y$ is given by
\begin{equation}
\rho=\frac{\text{Cov}(X,Y)}{\sigma_{X} \sigma_{Y}}=\sqrt{\frac{\sigma_D^2 + \sigma_A^2}{\sigma_D^2 + 2\sigma_A^2}}.
\end{equation}
Therefore, the joint probability density function of the bivariate normal random variables $(X,Y)$ is given by
\begin{equation}
\begin{aligned}
f_{X,Y}(x, y)
&= \frac{1}{2 \pi \sigma_{X} \sigma_{Y} \sqrt{1 - \rho^2}}
\exp \left(
-\frac{1}{2(1 - \rho^2)}
\left[
\frac{x^2}{\sigma_{X}^2}
- \frac{2\rho\, x y}{\sigma_{X} \sigma_{Y}}
+ \frac{y^2}{\sigma_{Y}^2}
\right]
\right) \\
&= \frac{1}{2 \pi \sigma_A \sqrt{\sigma_D^2 + \sigma_A^2}}
\exp \left(
-\frac{\sigma_D^2 + 2\sigma_A^2}{2\sigma_A^2}
\left[
\frac{x^2}{\sigma_D^2 + 2\sigma_A^2}
- \frac{2 x y}{\sigma_D^2 + 2\sigma_A^2}
+ \frac{y^2}{\sigma_D^2 + \sigma_A^2}
\right]
\right).
\end{aligned}
\end{equation}
Hence, the probability density function of $u_{P3}'$ is given by
\begin{equation}
\begin{aligned}
f(u_{P3}') 
&= \iint f_{X,Y}(x,y)\, \delta \big(u_{P3}' - x + R{\sqrt{\pi}}(y)\big) \, dx\, dy \\
&= \sum_{n \in \mathbb{Z}} \int_{(n - \frac{1}{2})\sqrt{\pi}}^{(n + \frac{1}{2})\sqrt{\pi}} f_{X,Y}\left( u_{P3}' + (y - n\sqrt{\pi}), y \right)  dy \\
&= \frac{1}{2}\sum_{n\in\mathbb{Z}} 
p[\sigma_A^2](u_{P3}'-n\sqrt{\pi})
\Bigg[
\operatorname{Erf}\!\left(
\frac{(1+2n)\sqrt{\pi/2}}{2\sqrt{\sigma_D^2+\sigma_A^2}}
\right)
- \operatorname{Erf}\!\left(
\frac{(2n-1)\sqrt{\pi/2}}{2\sqrt{\sigma_D^2+\sigma_A^2}}
\right)
\Bigg],
\end{aligned}
\end{equation}
where $p[\sigma^2](z) \equiv
\frac{1}{\sqrt{2\pi\sigma^2}}
\exp\!\left[-\frac{z^2}{2\sigma^2}\right]$. By a similar procedure, we can obtain the probability density functions of $v_{P3}'$, $u_T'$, and $v_T'$, which are found to follow identical distributions. This confirms that the P-Steane scheme with $(m=1, b=\sqrt{2})$ and the teleportation-based scheme achieve identical performance.

\end{widetext} 
\bibliography{ref_GKP}

\end{document}